\documentclass[acmsmall, anonymous=False, review=False]{acmart}

\usepackage{multirow}
\usepackage{booktabs}
\usepackage{tabularx}
\usepackage{float}
\newcommand{\blue}[1]{{\color{black} #1}}

\AtBeginDocument{%
  \providecommand\BibTeX{{%
    \normalfont B\kern-0.5em{\scshape i\kern-0.25em b}\kern-0.8em\TeX}}}

\setcopyright{acmlicensed}
\acmJournal{PACMHCI}
\acmYear{2022} 
\acmVolume{ } 
\acmNumber{CSCW} 
\begin{document}

\title[Privacy Research with Marginalized Groups]{Privacy Research with Marginalized Groups: \\ What We Know, What's Needed, and What's Next}

\author{Shruti Sannon}
\email{sannon@umich.edu}
\affiliation{%
  \institution{University of Michigan}
  \city{Ann Arbor}
  \state{MI}
  \country{USA}
}

\author{Andrea Forte}
\email{aforte@drexel.edu}
\affiliation{%
  \institution{Drexel University}
  \city{Philadelphia}
  \state{PA}
  \country{USA}
}


\begin{abstract} 

People who are marginalized experience disproportionate harms when their privacy is violated. Meeting their needs is vital for developing equitable and privacy-protective technologies. In response, research at the intersection of privacy and marginalization has acquired newfound urgency in the HCI and social computing community. In this literature review, we set out to understand how researchers have investigated this area of study. What topics have been examined, and how? What are the key findings and recommendations? And, crucially, where do we go from here? Based on a review of papers on privacy and marginalization published between 2010--2020 across HCI, Communication, and Privacy-focused venues, we make three main contributions: (1) we identify key themes in existing work and introduce the \textit{Privacy Responses and Costs} framework to describe the tensions around protecting privacy in marginalized contexts, (2) we identify understudied research topics (e.g., race) and other avenues for future work, and (3) we characterize trends in research practices, including the under-reporting of important methodological choices, and provide suggestions for establishing shared best practices for this growing research area.

\end{abstract}



\maketitle

\section{Introduction}

People who face marginalization in society---whose needs and experiences are overlooked and who have limited resources and power due to some facet of their identity ~\cite{cook2008marginalized}---can have unique privacy-related needs and behaviors that must be recognized by researchers and designers of technology. Marginalized groups can experience disproportionate harms when their privacy is violated. Consider the case of a person living with HIV, who may risk social stigma, discrimination, or the dissolution of relationships should their HIV status become publicly known without their consent \cite{greene2003privacy}, as compared to the impact of a health information leak for a person who does not have to navigate a marginalized and often stigmatized identity. In response to these elevated privacy threats, marginalized individuals have to navigate complicated decisions around identity management and disclosure, which extend to their technology use \cite{sannon2019really}. At the same time, marginalization can also constrain privacy-protective behaviors; for example, people in economically disadvantaged communities often have limited access to privacy literacy resources \cite{li2018privacy}. Understanding the new dimensions and consequences that privacy issues take on in marginalized contexts is vital to designing more equitable technologies that respect the privacy of all users rather than a select few. 

Given the need to represent marginalized voices in the design of technologies, and the potential privacy harms that stem from excluding them, there has been growing interest in this research area \blue{\cite{walker2019moving}}. Wang suggests that we are currently seeing a third wave in privacy research that centers what he terms as ``inclusive privacy'' that encompasses ``different human abilities, characteristics, needs, identities, and values,'' as compared to prior waves that focused on the technical and usability aspects of privacy \cite{wang2017third, wang2018inclusive}. Concurrently, privacy researchers in this space have begun to convene via panels \cite{das2020humans} and workshops, such as the 2017 CSCW workshop on privacy for vulnerable populations \cite{mcdonald2020privacy}, and the 2015-2022 SOUPS workshops on inclusive privacy and security (WIPS), pointing to a clear interest in this research area.  

What remains unclear is how the work being conducted by researchers in this area coheres together, the bounds of communal knowledge in this space, and what gaps still need to be filled. Literature reviews can be useful to assess and guide new, emerging areas of research \cite{dillahunt2017sharing}. Literature reviews have been used to this end in computing research, including at CSCW and CHI (e.g., \cite{dillahunt2017sharing, disalvo2010mapping}). Given the rapidly growing interest and research in this space, we see this as a key moment in time for a literature review that could help spur new research in this area while also providing an understanding of our cumulative knowledge and practices thus far. 

%What we did
To this end, we conducted a literature review of privacy research on marginalized populations, \blue{using a broad and open-ended definition of ``marginalized adults'' as adults facing any form of social exclusion or discrimination due to some facet of their identity \cite{hall1994marginalization, cook2008marginalized}, such as along the lines of race, sexual identity, gender identity, socioeconomic status, and immigration status, among other factors. We examined papers} published between 2010--2020 across a broad range of venues representing three inter-related disciplines: HCI and Social Computing (e.g., Proceedings of CSCW, CHI, and Ubicomp), Communication (e.g., Journal of Communication and New Media and Society), and Privacy-focused venues (e.g., Proceedings of the Privacy Enhancing Technologies Symposium). 

After manually reviewing 2,823 privacy-related papers that were published across these disciplines between 2010-2020, we identified 88 papers that focused on the intersection of privacy and marginalization. Based on an analysis of these 88 papers, we offer the following contributions:

\begin{itemize}
  \item \emph{Key Research Areas and Findings.}
  We describe the major contexts of marginalization studied and the kinds of findings and recommendations that papers in our dataset put forward. \blue{In doing so, we introduce the \textit{Privacy Responses and Costs} framework, which outlines 10 privacy responses people have to privacy threats, and the costs and consequences of these responses faced by marginalized groups.}
  \item \emph{Descriptive Norms for Research Practices.} We characterize the literature in terms of the practices that researchers report using to study privacy and marginalization. 
  \item \emph{Gaps and Implications.}
  Having surveyed the literature, we identify understudied areas and questions (e.g., race), lack of diversity in methods, and under-reporting of important methodological choices, and we make concrete recommendations for establishing shared best practices for this growing research area, such as designing protocols that minimize harm.
\end{itemize}

\section{Related Work}

Given our focus on marginalization, we begin by explaining what we mean when we use this term. Then, we discuss the use of literature reviews in computing research broadly and privacy research both within and outside of HCI and social computing (for brevity, we refer to this field as just ``HCI'' in what follows). We end by discussing the focus of the current review.

\subsection{What Do We Mean by Marginalization?}

Although the terms ``marginalization'' or ``marginalized'' are sometimes used in studies of technology use and privacy concerns, the term is commonly left undefined in HCI and social computing work. In our findings, we used a grounded approach to identify the ways that researchers engage and discuss marginalization in the privacy literature; however, to begin the task of identifying research about marginalization, we needed to define precisely what we mean when we use the term. 

%Moved up from the Methods section:
Marginalization is a complicated concept with varying definitions and dimensions. We drew from several sources to construct our frame for ``marginalized contexts.'' Marginalized populations are defined as ``persons who are peripheralized based on their identities, associations, experiences and environments'' \cite[p. 25]{hall1994marginalization}. As a result, these groups are ``excluded from mainstream social, economic, cultural, or political life'' \cite{cook2008marginalized}. People can be marginalized based on several factors, including race, disability, gender identity, sexual orientation, socioeconomic status, and immigration status. 

The impact of marginalization can be seen not only in the ways groups are peripheralized, but also in social responses to this exclusion. Marginalization has been acknowledged through legal protections around the world through laws prohibiting discrimination in housing, employment, and other areas of life (e.g., the Fair Housing Act in the U.S. \cite{USDOJ_2021}, \blue{the Racial Equality directive in the European Union \cite{eu-racial}, and the Rights of Persons with Disabilities Act in India \cite{india-disabilities-act})}. 

In this review, we focus on research about marginalized adults, many of whom face discrimination that these legal responses were set up to address. Although we were undoubtedly influenced by these readings and familiar a priori categories of marginalization commonly represented as legally protected classes, we conducted an open-ended analysis: if a paper described a group as being stigmatized or marginalized, or referred to privacy and marginalization more broadly, we opted to include it. As a result, our review includes a wide swath of marginalized identity characteristics, such as disability, sexual orientation, socioeconomic status, immigration status, and race, as well as experiences like sex work and human trafficking.

Although the studies we review in this paper are included because they study populations that have the characteristic of being ``marginalized,'' marginalization is not simply a feature of a population or individual, but a dynamic social process of exclusion that marshals the power of social norms, institutions, and interpersonal dynamics to render some people as privileged and others as inconsequential. As privacy researchers, we use this process orientation to understand privacy risks and violations as features of marginalization that change over time and in different contexts. As social computing scholars, we pay special attention to the role that technology design plays in initiating, continuing, accelerating, stalling, or obstructing processes of marginalization.

\subsection{Why do a Literature Review? Reviews in HCI and Social Computing} 

Literature reviews are not uncommon in HCI and social computing venues, and can provide insight into both the knowledge built by a research community as well as the practices used to construct this knowledge. First, literature reviews are a useful way to identify research trends and unearth new directions for research communities, and have been used to this effect to further research on the sharing economy \cite{dillahunt2017sharing}, HCI for development \cite{dell2016ins}, and sustainable HCI \cite{disalvo2010mapping}, among other areas.

Second, literature reviews can also help develop clarity around methods and direct best practices for a research community. To this effect, literature reviews have provided insight and instruction around the use of reliability measures \cite{mcdonald2019reliability}, the reporting of compensation \cite{pater2021standardizing}, the standards for sample size \cite{caine2016local}, and anonymization practices \cite{abbott2019local} in social computing research. 

Our goal is for this review to serve both of these purposes: (1) to understand the existing knowledge generated by researchers on privacy issues in marginalized contexts while pinpointing new areas of exploration, and (2) to understand the methods through which this knowledge generation has occurred with an eye towards guiding future research practices.

\subsection{Reviews of Privacy Research}

Literature reviews are also not new to privacy research, both within and outside the HCI research community. Some of these reviews have explored conceptual or theoretical approaches in privacy research, such as how design relates to privacy in HCI research \cite{wong2019bringing}, how researchers have theorized about the privacy paradox \cite{barth2017privacy}, and how privacy has been conceptualized in HCI \cite{barkhuus2012mismeasurement}. 

Alternatively, several privacy-related reviews have focused on the privacy considerations of specific technologies or contexts, such as eye-tracking \cite{katsini2020role}, cryptocurrencies \cite{herskind2020privacy}, Internet of Things (IoT) infrastructures \cite{alshohoumi2019systematic}, big data \cite{nelson2016security}, or electronic health record systems \cite{mahfuth2016systematic, rezaeibagha2015systematic}. These reviews have put forth several implications for privacy research, including identifying privacy threats that still need to be addressed \cite{herskind2020privacy}, and identifying a dearth of research on privacy protection in relation to the specific technologies under study \cite{katsini2020role, alshohoumi2019systematic}.

\subsection{Our Focus} %where we come in

It is clear that literature reviews can be useful in evaluating nascent computing fields and subfields and unearthing future directions for research communities. We searched the ACM and other databases to determine whether a review of privacy research on marginalized contexts had been conducted, and to our knowledge, no such review exists. Given both the importance of and growing interest in this subfield of privacy, we aim to provide a review that will tie existing work together, highlight key similarities and differences in the subfield's knowledge and research practices, and provide recommendations for future research. 

Having established the need for a review of research on privacy and marginalization, we drew on existing literature review practices to appropriately scope our inquiry. First, many literature reviews published in HCI and social computing venues (e.g., \cite{abbott2019local, mcdonald2019reliability}) %, including those on privacy (e.g., \cite{wong2019bringing}) 
focus on research published by the ACM. Since our primary audience is the HCI and CSCW community, we also chose to focus on HCI research published by the ACM. However, we also wanted to acknowledge the blurry nature of interdisciplinary boundaries. This led us to review a sample of publications from adjacent disciplines as well. Toward this end, we also examined papers published in a selection of high-impact Communication journals like \textit{New Media and Society} and Privacy-focused venues like \textit{IEEE Privacy \& Security}. 

Second, many social computing literature reviews are scoped by time period. We found considerable variation among the publication periods examined by reviews, ranging from three years \cite{mcdonald2019reliability} to ten years \cite{dillahunt2017sharing}. We scoped our review to papers published from 2010 through 2020 to capture a broad range of papers and measure activity in this research area over time. 

\blue{As discussed in Section 2.2, reviews can synthesize knowledge in a domain, identify areas ripe for exploration, and develop best practices for a research community.  Accordingly, our review is guided by a few overarching questions to achieve these goals.} The first set involves the breadth of research in this area and key findings: 
\begin{quote}
    What research has been done on privacy in marginalized contexts? What can we learn from this cumulative knowledge about how marginalization relates to privacy, and what are the opportunities for future research directions in this space?
\end{quote}

The second set of questions involves the practice of doing research on this topic: 
\begin{quote}
    How has this research been done? What can we learn in terms of best practices for conducting privacy research on marginalized contexts?
\end{quote}

\section{Method}

Our method for conducting the review was made up of four main stages: 1) collecting a corpus of privacy-related papers in multiple fields and venues, 2) filtering this dataset to only include papers that focus on both privacy and marginalization, 3) conducting a quantitative analysis to identify descriptive trends (e.g., common publication venues), and 4) conducting a qualitative thematic analysis to understand themes in the papers' approaches, findings, and recommendations. 

\begin{figure}[t]
\centering
\includegraphics[scale=0.6]{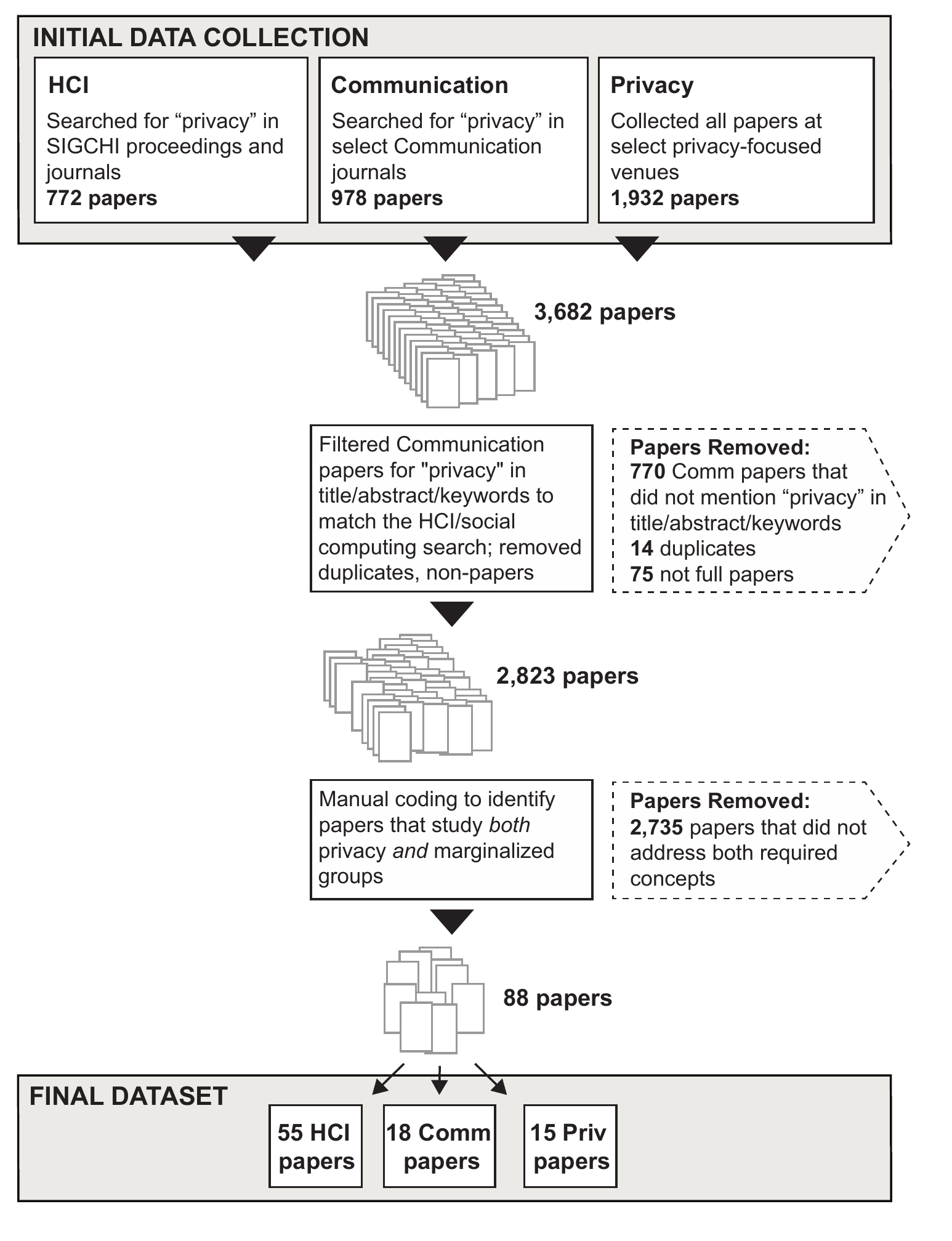}
\vspace{-5mm}
\caption{The Preferred Reporting Items for Systematic Reviews and Meta-Analysis (PRISMA) schematic of study flow, illustrating the data collection and filtering process of publications in selected venues published during 2010--2020}
\Description[A visual depiction of the Preferred Reporting Items for Systematic Reviews and Meta-Analysis (PRISMA) schematic of study flow. All of the data presented in this graph is also given in the text, describing how we collected and filtered the data at each stage.]{
}
\label{fig:Flowchart}
\end{figure}

In what follows, we explain the steps we took at each stage of this process. We also present an overview of our method using the Preferred Reporting Items for Systematic Reviews and Meta-Analysis (PRISMA) schematic of study flow \cite{moher2009preferred} in  Figure~\ref{fig:Flowchart}.

\subsection{Data Collection}

To reflect the diversity of research on privacy, we sampled papers from three distinct but interdisciplinary and overlapping fields: HCI, Communication, and Privacy-focused venues. \blue{Our goal was to review a broad set of papers that represent research in these fields, rather than attempting to collect the complete set of papers published on privacy and marginalization. }

Data collection took place in 2021. Per guidelines for effective literature reviews \cite{khan2003five}, we developed selection criteria to identify relevant papers for inclusion in the review: (1) full research articles (i.e., not posters, abstracts, panels, and so forth), (2) published between 2010 and 2020, (3) with a focus on both privacy and marginalization. In this section, we describe our process for identifying search terms for the review, and then detail our data collection procedures for each of the three fields.

\subsubsection{Identifying Search Terms}

We began by exploring search terms for identifying relevant papers. Since our aim was to identify papers that were both about privacy and marginalization, we considered using search terms for both of these foci. However, we quickly discovered that many relevant research papers on marginalized contexts do not use common keywords (such as ``marginalization'' or ``vulnerable'') and would thus be excluded from our dataset. While we considered developing a list of marginalized contexts to use as a priori search terms (e.g., ``disability''), we decided that this would place artificial limits on the breadth of our data. 

Thus, we decided to take a more expansive (albeit labor-intensive) approach: to collect all privacy-related papers published in our selected venues to arrive at a dataset of privacy research between 2010--2020, and then manually sort through these papers using our selection criteria to identify papers that also focused on marginalized contexts. The venues we selected to gain a broad overview of HCI, Communication, and Privacy venues are listed in Table~\ref{tab:venues}. 
To identify privacy-related papers, we searched for the term ``privacy'' in papers' titles, abstracts, and/or keywords. We considered using a more extensive set of keywords related to privacy (e.g., surveillance, tracking, disclosure), but found that these terms were often overly broad or narrow, and thus introduced significant noise in the dataset (e.g., ``tracking'' brought up numerous results about eye sensor tracking that were not related to privacy). Since we were also searching fields outside of HCI where these terms could take on conflicting dimensions, we decided to use privacy as the sole keyword.

\begin{table}[b]
\begin{tabular}{@{}ll@{}}
\toprule
Venues & Journals and Proceedings \\ \midrule
HCI & CSCW, CHI, DIS, MobileHCI, PACM CSCW, PACM GROUP, PACM IMWUT \\
Communication & CR, JOC, IJOC, JCMC, NM\&S, SM+S \\
Privacy & SOUPS, PoPETS, IEEE S\&P, USENIX Security \\ \bottomrule
\end{tabular}
\caption{Venues represented in the dataset comprising 8 conference proceedings and 9 journals from 2010--2020. Note that CSCW was a proceedings through 2016 and a journal thereafter.}
\Description[Table]{A table that lists the 8 conferences and 9 journals in the dataset; the list is also provided in the text of the paper.}
\label{tab:venues}
\end{table}

\subsubsection{HCI}

We searched the ACM Digital Library (ACM-DL) for papers with the term ``privacy'' in either the title, abstract, and/or keywords. We used the ACM-DL search options to restrict the results to (1) papers published between 2010--2020, and (2) papers classified as ``Research article'' (thus excluding posters, workshops, abstracts, and panels). 

We then selected the ``sponsored by SIGCHI'' option to identify papers published in the proceedings of HCI conferences (e.g., CSCW, CHI, DIS, etc.). This search produced 623 papers. We also collected search results for the following journals: the Proceedings of the ACM on Human-computer Interaction (PACM) (62 papers), Proceedings of the ACM on Interactive, Mobile, Wearable and Ubiquitous Technologies (IMWUT) (75 papers), and Transactions of Computer-Human Interaction (TOCHI) (12 papers). 

We downloaded a dataset composed of 772 HCI papers based on the above procedure. We cleaned this dataset by removing 14 duplicates and 75 non-research articles (e.g., posters and doctoral consortium abstracts that were incorrectly tagged in the ACM-DL). Our HCI dataset after data cleaning contained 683 privacy-related papers. 

\subsubsection{Communication}

There is no central database for Communication papers, and literature reviews on communication and technology typically examine a predetermined selection of journals (e.g., \cite{evans2017explicating}). Following this approach, we selected a sample of top journals in Communication that publish papers on technology: Journal of Communication (JOC), Journal of Computer-Mediated Communication (JCMC), Communication Research (CR), International Journal of Communication (IJOC), New Media and Society (NMS), and Social Media + Society (SM+S).   

We searched for ``privacy'' on each individual journal's website, filtering for full research articles published during 2010--2020. This resulted in a total of 978 papers (JOC = 58, JCMC = 155, CR = 63, IJOC = 79, NMS = 380, and SM+S = 243). However, some journals' search interfaces did not allow us to exclude full-text searches (unlike our ACM search), and thus many papers were irrelevant to privacy as a research topic (e.g., papers that mentioned privacy once in the Methods section when describing a consent process). Thus, to stay consistent with our ACM search, we manually filtered the Communication dataset to only include papers that contained privacy in their titles, abstracts, and/or keywords (when available). After this filtering process, the Communication dataset consisted of 208 privacy-related papers (JOC = 8, JCMC = 13, CR = 4, IJOC = 79, NMS = 58, SM+S = 46).

\subsubsection{Privacy-focused venues}

We sampled four Privacy venues: the USENIX Symposium on Usable Privacy and Security (SOUPS), IEEE Security and Privacy (S\&P), Proceedings of the Privacy Enhancing Technologies Symposium (PoPETS), and USENIX Security. Since these venues are focused on privacy, we did not have to run a search to identify privacy-related papers. Instead, we included the entire corpus of 2010--2020 papers published at SOUPS, IEEE S\&P and USENIX Security in our dataset; PoPETS began its proceedings in 2015, and we included all PoPETS papers from 2015-2020 in the dataset. The dataset from these venues consisted of 1,932 papers (SOUPS = 232, PoPETS = 306, USENIX Security = 813, IEEE S\&P = 581).

\subsection{Data Filtering}

Once we had a corpus of 2,823 privacy-related papers from HCI, Communication, and Privacy-focused venues, we manually coded this dataset to identify papers that focused both on privacy and marginalization. We coded each paper for inclusion/exclusion according to the following codes: (1) relevant (the paper is about privacy in the context of marginalization), (2) not relevant (the paper does not cover both privacy \textit{and} marginalization), and (3) unclear. We coded each paper based on its title and abstract, only referring to the full text of the paper if the coding decision was unclear. 

\blue{We operationalized marginalization through an iterative process of reading and discussion---in addition to reading in the areas of HCI, Communication, and Privacy, we investigated the history of the term and read widely in journals from other fields, books, and websites of governmental agencies and NGOs. Definitions of marginalization in these readings were not homogeneous and sparked lively discussions in our research group about inclusion criteria. In cases where a group's marginalized status was unclear, we chose to only include papers that used language to plainly situate such populations as marginalized.}

To ensure a clear and consistent interpretation of the concepts ``privacy'' and ``marginalization'', we used an iterative process of discussion and testing. Two researchers independently coded a subset of 10 papers each for relevance and discussed their initial impressions. Then, they coded 25 more papers and met to discuss and reconcile disagreements; at this point, their inter-rater reliability was good (Cohen’s $\kappa$ = 0.74). Following this discussion, they coded 25 additional papers; inter-rater reliability for this round of coding was strong (Cohen’s $\kappa$ = 0.83). Having established close agreement over the coding scheme, one coder then independently coded the remaining papers in the dataset for relevance. Out of an abundance of caution, 61 papers were coded as ``unclear'' and flagged for further examination; through discussion, 13 of these papers were re-coded as relevant for inclusion in the final dataset.  

Ultimately, 88 of the 2,823 privacy-related papers were coded as relevant for inclusion in the final dataset (HCI = 55, Communication = 18, Privacy = 15). All 88 papers were downloaded from their respective repositories for analysis. \blue{A list of the 88 papers is available as a supplementary file to this paper on the ACM digital library}.

\subsection{Data Analysis}

During analysis, both authors read papers in the dataset; between us, we read all 88 papers. We conducted two main analyses: a quantitative analysis to identify descriptive trends across papers, and a qualitative analysis to identify themes in papers' findings and practices, as detailed below.

\subsubsection{Quantitative Analysis}

We extracted descriptive data about each paper and compiled these into a spreadsheet. These data included each paper's publication year, publication venue, methods, marginalized population/context, and technological focus, if any. We also noted whether each paper discussed ethics, compensation, and positionality; we extracted the content of these statements---when present---into the spreadsheet. We used these data to calculate descriptive statistics about the dataset (e.g., how many papers were published at CSCW, how many focused on LGBTQ+ individuals, how many provided information about compensating participants). 

\subsubsection{Qualitative Analysis}

As we read through each paper, we created open codes to identify concepts and themes that cut across the dataset. We focused on study rationales, findings, implications, and recommendations of each paper, as well as how authors discussed the relationship between technology, privacy, and marginalization. We met regularly to discuss patterns across papers. For example, we found that several papers highlighted key privacy-related tensions in relation to marginalization (such as the need for social support versus the need for privacy); through discussion, we categorized these as four key tensions that we present in Section 7.2. 

The primary goal of the qualitative analysis was to inductively generate concepts and themes, not to consistently identify examples of pre-defined concepts; as such, we did not calculate inter-rater reliability but instead used ongoing, iterative discussion to achieve consensus, aligned with best practices described in  \citet{mcdonald2019reliability}. This is in line with several other reviews that focus on surfacing themes in research areas (e.g., \cite{disalvo2010mapping, wong2019bringing}).

\section{An Overview of the Dataset}

We begin by providing a snapshot of the work done at the intersection of privacy and marginalization. We describe when and where papers in our dataset were published, and then identify the foci of these papers, both in terms of the types of marginalization contexts and the types of technologies they focus on. To describe our findings, we use the following notation for each main type of venue: HCI (SIGCHI venues), C (Communication), and P (Privacy-focused venues). 

\subsection{Publication Venues and Years}

\begin{figure}[t]
\centering
\includegraphics[scale=0.6]{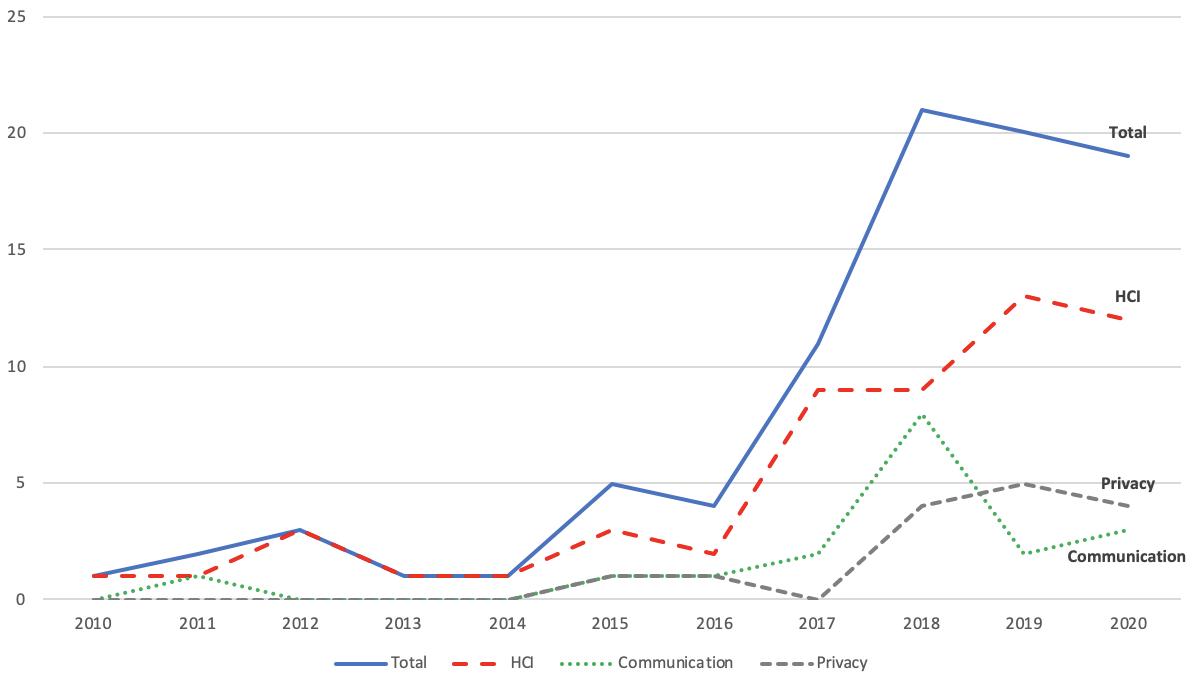}
\vspace{-5mm}
\caption{Publication counts from 2010 to 2020 in HCI, Communication, Privacy-focused venues, as well as the total number of papers across venues over time}
\Description[A line graph]{A line graph showing a steep increase in papers from 2017 onwards}
\label{fig:years}
\end{figure}

%where
Our final dataset represented 88 papers from 17 journals or conference proceedings. The majority of these papers were in HCI (55, 63\%), followed by Communication (18, 20\%) and Privacy-focused venues (15, 17\%). Figure~\ref{fig:years} illustrates the number of papers published between 2010--2020 by venue.

Within HCI, the majority of the 55 papers were published at CHI (24, 44\%) and CSCW (including both the conference proceedings and the PACM journal, 24, 44\%), suggesting that CHI and CSCW are the key venues that HCI researchers target for work on privacy and marginalization and that these two communities have an orientation that makes them receptive to the work. A few were published at other HCI venues: DIS (4, 7\%), PACM IMWUT, GROUP, and MobileHCI (1 paper each). 

In Communication, a little over half of the 18 papers were from the International Journal of Communication (10, 56\%), followed by New Media and Society (4, 22\%) and Social Media + Society (2, 11\%). Journal of Communication and Communication Research included one paper each. 

In Privacy-focused venues, most of the 15 papers were in SOUPS (6, 40\%) and USENIX Security (5, 33\%), followed by PoPETS and IEEE S\&P (2, 13.33\% each). 

%when
Overall, the vast majority (71, 78\%) of the 88 papers were published between 2017 and 2020, bolstering our view that this is a growing area of interest among researchers across HCI/Social Computing, Communication, and Privacy. 

\subsection{Types of Marginalization Contexts} %who

\begin{figure}[b]
\centering
\includegraphics[scale=0.5]{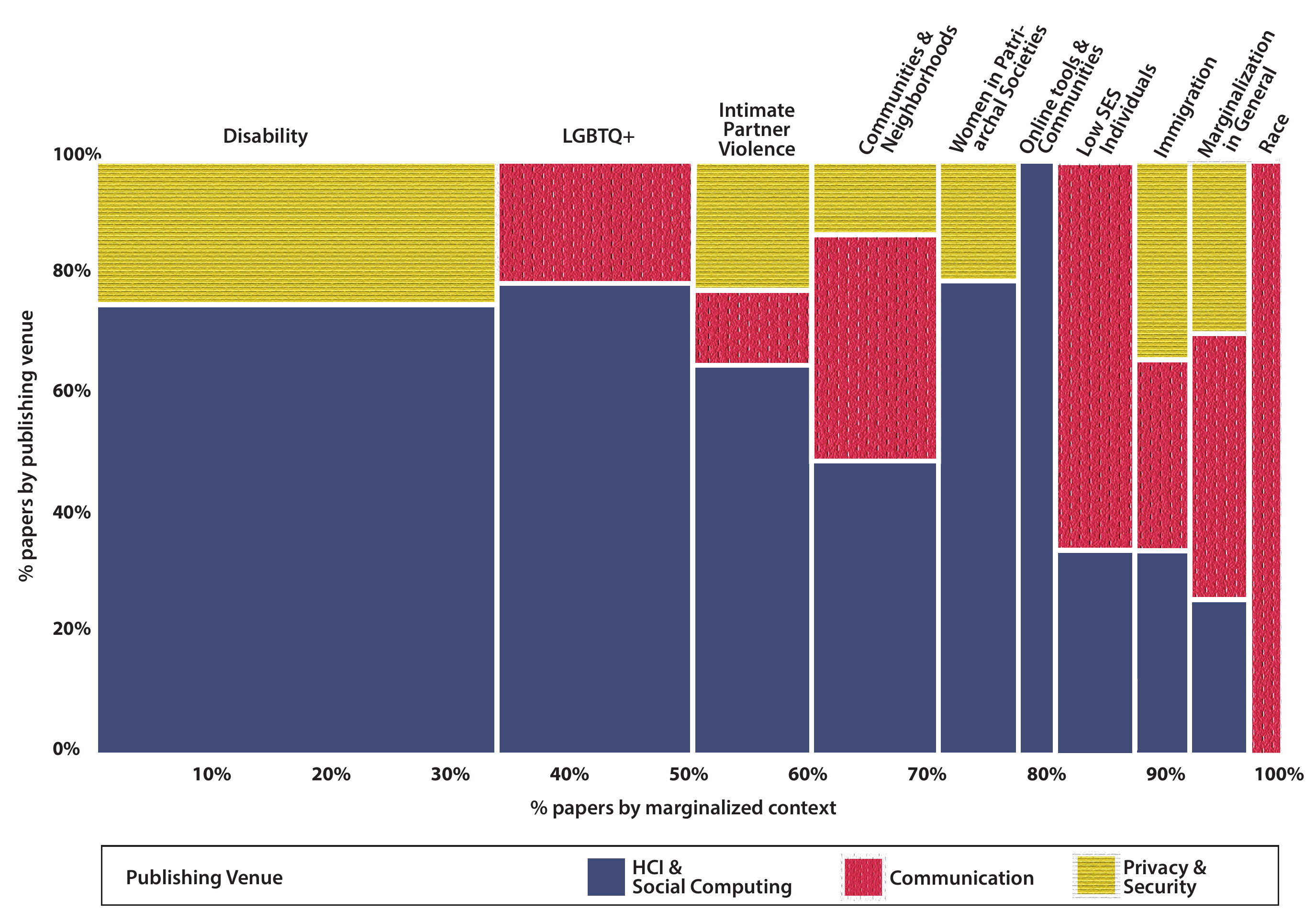}
\caption{Papers by publication venue and marginalization context. The width of each bar represents the percentage of papers in each marginalization context across all publication venues. The height of each color bar represents the percentage of papers in each publishing venue. Graph includes contexts represented by at least 2 papers.}
\Description[A mosaic graph]{A mosaic graph showing that the majority of papers across contexts in the dataset are in HCI; Disability and LGBTQ+ issues make up half of all contexts}
\label{fig:mosaic}
\end{figure}

We found that papers could be broadly categorized as focusing on the following types of marginalization contexts: (1) individuals and identities; (2) physical spaces and communities; (3) online spaces, tools, and communities; or (4) marginalization in general. We discuss each of these types below. Figure~\ref{fig:mosaic} presents a breakdown of paper distributions by context and publishing venue.

\subsubsection{Individuals and Identities} 
Most papers (72, 82\%) focused on specific marginalized identities (HCI = 46, C = 13, P = 13). Disability was the most popular topic (28, 32\%) in the dataset, comprising more than a third of all HCI papers (21) and approximately half of papers at Privacy-focused venues (7). There were no Communication papers in this area. The most common disability topics included dementia and age-related cognitive impairments (HCI = 8, P = 4), visual impairments (HCI = 2, P = 4), and HIV (HCI = 5). In contrast, only three papers explored mental health conditions. 
 
The second most common focus was individuals who identify as LGBTQ+ (14, 16\%); within this category, most (8) papers focused on trans individuals. Several papers examined intimate partner abuse (9, 10\%), low-income individuals (6, 7\%), and the risks faced by women in patriarchal South Asian contexts (5, 6\%). There were also several papers that focused on contexts that often (but not exclusively) involve women, including sex work (1), sexual assault (1) and human trafficking (1), as well as women transitioning from incarceration (1).
    
Only two papers examined race and ethnicity as their primary focus; both in Communication. A few papers studied groups that were also often racial or ethnic minorities: undocumented immigrants (2), politically marginalized minorities (1), and refugees (1). 

\subsubsection{Physical Spaces and Communities}

Rather than focusing on individuals, some studies (10, 11\%) focused on marginalization at the level of communities (HCI = 6, C = 3, P = 1). Most of these papers (8) examined communities that are economically disadvantaged, while two examined other factors of marginalization: crime, income, and racial segregation across neighborhoods \cite{israni2017snitches} %(p20)
and communities that are ``geographically and culturally marginal'' \cite{jack2019privacy}.
%(p41)

\subsubsection{Online Spaces, Tools, and Communities}

Two HCI papers in the dataset focused on tools that marginalized groups use to protect their privacy online or communities in which marginalized people participate: Tor, open-source software that enables anonymous communication \cite{forte2017privacy}, and online fandom communities \cite{dym2020social}.

\subsubsection{Marginalization in General}

Four papers did not focus on a particular context of marginalization: One non-empirical paper theorized about how to integrate vulnerability and marginalization in privacy theories and research \cite{mcdonald2020politics}. Two explored how sensitive data about people can be identified through their digital traces \cite{maris2020tracking, cabanas2018unveiling}, and how these privacy violations can be particularly concerning for populations with stigmatized or marginalized attributes. One paper examined how the digital divide impacts marginalized groups more broadly \cite{gangadharan2017downside}.

\subsection{Types of Technologies} 

In addition to identifying the marginalization context of papers, we identified the types of technologies they focused on, if any. Approximately one third of papers (31, 35\%) did not focus on a specific technology but, rather, explored technology use more broadly. This approach appeared in all three fields (HCI = 20, C = 5, P = 6). 

In terms of studies that focused on specific technologies, the most common focus was social media (23, 26\%), particularly in HCI (HCI = 18, C = 3, P = 2); these included papers that examined a broad range of social media platforms or individual platforms, as well as different types of social media (e.g., online forums, dating apps, social networking sites, etc.). 

Mobile phones were the second most common technology (8, 9\%), followed by location-tracking technologies (5, 6\%); the rest of the studies focused on many different types of technologies, including assistive technologies, automatic gender recognition algorithms, smart homes, and national identification systems. In a departure from the rest of the dataset, two Communication papers did not focus on technology or technology use at all, instead theorizing about the relationship between marginalization and privacy.

\section{Why Study Privacy and Marginalization?} % "getting started"

To study a phenomenon, it must be problematized in such a way that its importance is clear to the members of a research community. How researchers frame marginalization as a compelling area for privacy research not only sheds light on the research topic, but also on the scholarly community. In this section, we explore the rationales motivating work in this space.%; these rationales were not mutually exclusive. 

\subsection{Disproportionate risks and challenges}

In many cases, studies problematized how marginalized groups may need to use technology---particularly in service of some identity-related goal---and simultaneously face elevated privacy risks from doing so. For example, survivors of intimate partner violence who shared custody of a child with an abusive ex-partner found that while electronic communication enabled remote contact between the child and ex-partner, it also opened up the possibility for intrusive access, such as through a live video feed of their home~\cite{freed2017digital}. 

Several studies were motivated by the potential ramifications of privacy violations and their disproportionate impact on marginalized groups. Maris et al. observe that leaked URL data can disclose sexual interests, and note that ``Those most likely to be impacted by online sexual privacy violations are traditionally marginalized and vulnerable communities, especially women, people of color, LGBTQ and other marginalized gender/sexuality communities'' \cite{maris2020tracking}. Data leaks about sexual preferences that are aligned with normative standards may prove embarrassing, but are not likely to lead to harassment or discrimination. However, such data leaked about sexual interests that are perceived as non-normative or stigmatized may lead to substantial harm.

While marginalized groups use a range of strategies to protect themselves from privacy risks, to resist technology, and to assert their agency \cite{hamby2018privacy}, they can also face disproportionate challenges in protecting their privacy due to disparities in digital literacy, skills, technological access as well as linguistic and cultural barriers (e.g., \cite{srinivasan2018privacy, ahmed2015privacy}). 

\subsection{Technological and societal exclusion}

Researchers also cited technological and societal exclusion as reasons to study marginalized groups. Marginalized groups' access to resources and opportunities are often bounded by their access to (and willingness to use) technologies that can violate their privacy. Technologies can potentially exacerbate social exclusion because of the privacy risks they pose, and in this way, they can reify or accelerate the process of marginalization.  

Some papers discussed limitations in various types of technologies, including phones and facial recognition software, with an emphasis on the fact that technology is often designed without consideration for the needs of marginalized groups. Ahmed et al. explain that for disabled persons, ``there are also significant differences [in privacy concerns], due in part to their disability but also because of systems that were not designed with them in mind'' \cite{ahmed2015privacy}.

Exclusion from the technology design process can create a host of privacy-related challenges for marginalized groups that warrant examination. For example, Dosono et al. uncover several barriers faced by people with visual impairments in using authentication to access websites \cite{dosono2015m}.   

Other papers called out structural challenges that marginalized groups face in society that complicate experiences in realms like education and employment that were not always tied to technology use; for example, in the case of children of undocumented immigrants to the United States, disclosures about immigrant status may be necessary for youth seeking support both online and offline, but carry the oppressive threat of deportation \cite{kam2020latinx}.

\section{How Do Researchers Study Privacy and Marginalization?} 
 
The majority of papers (97\%) were empirical research (HCI = 54, C = 16, P = 15). Qualitative methods were most common across the board (HCI = 49, C = 12, P = 10), making up 81\% of all papers. Here, we provide a field-specific breakdown of methods used in this space.

%HCI deep dive para
In HCI, several studies used multiple methods (e.g., interviews as well as focus groups or a survey); here, we present counts for each method used (i.e., some papers were counted more than once). Semi-structured interviews were the most used qualitative research method (45), followed by focus groups (9), participant observation (6); some papers also drew on textual content analyses (3), participatory action research (1), and ethnographic action research (1). Seven HCI studies used quantitative methods, including a survey (1), experiment (1), computational techniques (2), system evaluation (2), and quantified measures to supplement an interview study (1). In addition, seven HCI papers involved the design, development, and/or evaluation of prototypes or systems, such as a sound awareness system for deaf people \cite{jain2020homesound}, a  monitoring system for dementia care \cite{wan2014addressing}, and a phone-based broadcast system to reach urban sex workers \cite{sambasivan2011designing}. 

Communication studies in this space were primarily qualitative, using interviews (10), observational fieldwork (2), ethnographic methods (2), and/or focus groups (1). The four remaining empirical Communication studies were quantitative, comprising of 3 surveys and a quantitative content analysis. 

Following the trend in HCI and Communication, all of the qualitative studies published in Privacy-focused venues drew on interviews (9). Computational methods were a little more common in Privacy-focused venues (4), such as natural language processing. 

Three papers in the dataset were theoretical pieces that were not based on empirical research (HCI = 1, C = 2) (e.g., \cite{reichel2017race}). 

In what follows, we discuss various aspects of researchers' methodological approaches and research practices, in terms of the types of lenses and frameworks they drew on, how they engage with marginalized groups, the ethical considerations of their research, decisions around compensation, and issues of positionality. 

\subsection{What lenses and frameworks do they draw on?}

A minority of papers explicitly employed a critical lens or framework in their research approach. Critical lenses are those that consciously reflect on how social norms and cultures create systems of oppression. In HCI, six papers used some form of critical lens, such as a feminist, queer-Marxist, or intersectional approach (e.g., \cite{clarke2013digital, sambasivan2019they}). In Communication, one paper provided a critical review of how privacy discourse has been historically used to further disenfranchise marginalized individuals along class and racial lines \cite{reichel2017race}. No studies in Privacy-focused venues employed these approaches.

\subsection{How do they engage with marginalized groups?} 

There is a growing call to engage marginalized groups directly in research involving them, and to empower them by providing opportunities to co-develop and shape research activities, findings, and recommendations. However, building rapport, establishing equitable collaborations, and navigating ethical issues like establishing appropriate consent processes and providing appropriate compensation for research participants means that engagement can be a complex issue for researchers. 

We examined the empirical studies in our dataset to identify how many engaged with marginalized groups as study participants. We found that participation or direct input from marginalized groups 
%(X, X\%) <- to get this, need to remove n/a like content analyses and theory papers and it's just messy..
was fairly common across fields  \blue{(HCI = 44, 80\%; C = 15, 83\%; P = 11, 73\%)}. %C = 15/16 (two theory n/a), P = 11/15 (4 n/a)

In many studies, participants from marginalized groups took part in interviews, surveys, and co-design sessions, among other research activities. In some cases, researchers also interviewed other stakeholders, such as familial caregivers and professional care workers. In three studies, the researchers explicitly discussed difficulties in recruiting their population of interest due to cognitive impairments that compromised both the consent process and the nature of responses; in these cases, these difficulties were included in the researchers' rationale to interview other stakeholders instead of the marginalized group themselves. 

\subsection{How do researchers discuss the ethical considerations of this work?} 

Questions of research ethics have become increasingly complicated in HCI research given the wide range of methods at researchers' disposal \cite{bruckman2014research}. At the same time, research ethics are a crucial consideration in studies involving marginalized groups. Research on sensitive topics can also engender challenges around ``situational ethics''---ethical considerations that emerge in the process of conducting research that depart from formal procedural ethics requirements \cite{munteanu2015situational}. Munteanu et al. provide examples of situational ethics, such as discovering that some participants are not comfortable with sharing in focus groups, and altering the method to one-on-one interviews instead to respect participants' needs \cite{munteanu2015situational}. 
In this section, we explore reporting practices about ethical considerations.

\begin{table}[b]
\begin{tabular}{@{}lcccc@{}}
\toprule
 & HCI & Communication & Privacy & Total \\ \midrule
Includes ethics discussion & 30 & 3 & 10 & 43 \\
No ethics discussion & 25 & 15 & 5 & 45 \\ \bottomrule
\end{tabular}
\caption{Papers that discuss ethical considerations}
\Description[Table]{A table that shows that about half of the papers in the dataset include ethical considerations; all numbers are also provided in the text of the paper.}
\label{tab:ethics}
\end{table}

As shown in Table~\ref{tab:ethics}, papers varied widely in whether and how they discussed ethical considerations. Almost half (49\%) discussed research ethics in some form---either in a dedicated section or integrated within a broader Methods section. However, this choice varied sharply by field, with most computing papers including such discussions (HCI = 30, 55\%, P= 10, 67\%) in contrast to Communication (C = 3, 17\%).  

Research ethics discussions ranged from cursory to extensively detailed. Researchers most commonly reported anonymizing data (e.g., by removing participant identifiers) and obtaining consent from participants; papers that had short ethics discussions invariably only reported these procedures. Some papers reported more extensively on planned or emergent ethics-related procedures. Although some procedures were uncommon, we discuss some key themes around privacy, trust, and support to illustrate the breadth of approaches to ethics in this research space. 

\subsubsection{Protecting Privacy}

Some papers went beyond anonymizing participant identifiers to protect participant privacy, such as by paraphrasing quotes taken from online data sources to reduce their searchability \cite{johnson2020roles} or not recording audio during interviews \cite{sambasivan2011designing}. One study took the uncommon approach of not asking for demographic information from participants, instead choosing to use existing demographic data from participants' neighborhoods as a proxy \cite{vitak2018knew}. 

\subsubsection{Building Trust}

Some studies involved developing trust with communities and/or community allies and soliciting their guidance during the research process. Some researchers reported developing relationships with community members over significant periods of time by attending gatherings and volunteering (e.g. \cite{hayes2019cooperative}). %add more 
Others consulted with community allies or non-profit organizations to determine appropriate compensation \cite{freed2017digital} and how to recruit their participants in a privacy-protective way \cite{guberek2018keeping}. 

Another study used member checking practices, where the researchers shared a draft of their paper with participants for feedback to ensure they had accurately represented their experiences and to correct any misunderstandings \cite{hayes2019cooperative}. 

\subsubsection{Prioritizing Support}

In some studies, researchers described how they prioritized the dignity of participants and supported them during the research process, particularly in the case of interviews about sensitive experiences. These research choices included stopping recording interviews when participants became emotional \cite{sambasivan2018privacy}, having a therapist present for interviews with survivors of intimate partner abuse \cite{leitao2019anticipating}, and appointing same-gender interviewers for women who have faced abuse \cite{sambasivan2019they}. Beyond participants, some papers discussed the steps taken to also support the research teams themselves because of exposure to difficult interview content \cite{havron2019clinical}.

\subsection{How do researchers navigate compensation with marginalized participants?}

Carefully navigating participant compensation is particularly important when studying marginalized groups, many of whom may be economically disenfranchised or otherwise vulnerable. While providing participants with compensation recognizes and respects their time and effort, financial incentives may also coerce people into participating in the research out of economic need \cite{ensign2003ethical}. Thus, whether and how to compensate participants requires careful consideration and review by institutional ethics boards on a case-by-case basis \cite{pater2021standardizing}. Reporting and explaining decisions around compensation can help readers evaluate the research ethics of any given study (e.g., the potential for undue influence) and its methodology (e.g., the impact of compensation on recruitment) \cite{klitzman2007reporting}.

In our dataset, fifteen papers did not involve marginalized populations directly (e.g., content analyses), and thus the question of compensation was irrelevant (HCI = 7, C = 4, P = 4). The remaining 73 papers involved some form of research activity with participants, and we examined these to understand trends and rationales around compensation; an overview is in Table~\ref{tab:compensation}.

Overall, 52\% (38) of the 73 user studies reported that participants were monetarily compensated (HCI = 25, C = 5, P = 8). Compensation ranged from \$5 to \$100 USD in cash, gift cards, or vouchers, and was typically provided for interviews. In a few studies, researchers determined appropriate compensation by consulting with expert organizations and communities (e.g., \cite{freed2017digital}), which can be a useful strategy when studying marginalized contexts \cite{ensign2003ethical}. In contrast, 41\% (30) of the 73 papers provided no information about compensation (HCI = 21, C = 6, P = 3). In addition, two HCI papers stated that they did not compensate participants, and three Communication papers stated that the researchers provided snacks and small gifts in lieu of financial compensation. 

\begin{table}[t]
%\small
\begin{tabular}{@{}lcccc@{}}
\toprule
 & HCI & Communication & Privacy & Total \\ \midrule
Provided monetary compensation & 25 & 5 & 8 & 38 \\
Provided small gifts or snacks & 0 & 3 & 0 & 3 \\
Explicitly did not provide compensation & 2 & 0 & 0 & 2 \\
Did not report compensation & 21 & 6 & 3 & 30 \\ \bottomrule
\end{tabular}
\caption{Compensation practices for the 73 studies involving participants}
\Description[Table]{A table that shows that 38 of 73 user studies provide compensation and 30 do not report compensation; all numbers are also provided in the text of the paper.}
\label{tab:compensation}
\end{table}

\subsection{How do researchers navigate positionality?}

Researchers' positionalities---their positions in society based on identity factors such as class, gender, and race, among others---invariably influence the research process in several ways, from the types of questions that are asked to how the work is completed \cite{england1994getting, parson2019considering}. Considering positionality is particularly important when conducting research with marginalized groups to ensure the research does not perpetuate the same marginalization that researchers seek to understand \cite{parson2019considering}. Positionality can also have implications for researchers' well-being. Insider research---when researchers belong to the same group as those being studied---in marginalized contexts can pose emotional risks for researchers, making self-care a critical part of such research \cite{shaw2020ethics}. 

Thus, we examined the papers in our dataset to identify whether and how researchers were engaging with issues of positionality. Overall, positionality statements were uncommon, with only 23\% of papers reporting positionality across the dataset (HCI = 14, C = 6, P = 0). Positionality statements typically involved one or more coauthors disclosing that they identified as belonging to the marginalized population that was the focus of the study. It was comparatively rare for authors to use these statements to discuss their relative privilege or non-membership in the groups being studied. Overall, positionality statements were brief and served to communicate authors' identities (either marginalized or not), and most papers did not discuss how these positionalities might inform or impact the research. Some studies in Communication used a less common practice that is worth noting. In these studies, when the authors did not belong to the population being studied, they enlisted population members to help inform the research process in terms of recruitment, feedback on questions, interviewing, and building rapport with the community. %(Paper SM+S1 is an example of a good methods section)

\section{What are key themes in the findings?}

In what follows, we discuss broad themes that reflect the ways in which marginalization, technology, and privacy were interrelated in the corpus of papers. \blue{ Central to this analysis are the many tensions that arise when people choose responses to perceived privacy threats, the high-level tradeoffs involved in these choices, and the specific costs and consequences of different responses to threats.}

\begin{table}
\small
\begin{tabular}{p{2.7cm}p{2.7cm}p{7.7cm}}
\hline \hline
\textbf{Privacy Response} & \textbf{Cost/Consequence} & \textbf{Select Examples from the Dataset} \\ 
\hline \hline
\textbf{Apathy}\newline i.e. lack of response & \textbullet\ Exposure to risks & Undocumented immigrants felt  government surveillance is inescapable, leading to inaction \cite{guberek2018keeping}; women transitioning from incarceration felt they have ``nothing to lose'' \cite{seo2020returning}. \\
\hline
\textbf{Non-use}\newline e.g. not using a technology, deleting an account & 
\textbullet\ Opportunity loss\newline\textbullet\ Exclusion\newline \textbullet\ Silencing \newline\textbullet\ Isolation & Economically disadvantaged populations lose opportunities due to non-use of technologies \cite{vitak2018knew}. \\
\hline
\textbf{Withholding \newline disclosure}\newline e.g. self-censorship, information removal & 
\textbullet\ Restricts \newline self-expression \newline\textbullet\ Silencing  & Low-SES youth \cite{marwick2017nobody}, marginalized Cambodians \cite{jack2019privacy}, and political refugees in the U.S. \cite{simko2018computer} self-censored to avoid conflict and danger but were further silenced by doing so. Men who have sex with men may not disclose HIV status on dating apps but can inadvertently signal positive status \cite{warner2018privacy}. \\
\hline
\textbf{Controlling \newline disclosure}\newline e.g. compartmentalizing identity,  multiple accounts, privacy controls, segmenting audiences &\textbullet\ Restricts \newline self-expression\newline\textbullet\ Labor-intensive\newline\textbullet\ Social cost \newline\textbullet\ Financial cost & Trans men crowdfunding top surgery used privacy controls to limit audiences \cite{fritz2018privacy}, young Azerbaijanis maintained multiple accounts for political activism \cite{pearce2018privacy}, and LGBTQ+ social media users managed identities across platforms \cite{devito2018too}. Disclosure controls require extensive labor and restrict self-expression \cite{blackwell2016lgbt}. Complex privacy controls can be costly to access \cite{reichel2020have} and can be used incorrectly due to accessibility issues \cite{ahmed2015privacy}. \\
\hline
\textbf{Privacy lies \cite{sannon2018privacy}} \newline i.e. providing false information  &\textbullet\ Cognitive burden\newline\textbullet\ Social/legal repercussions & Rural Appalachians provided false information as a form of vigilanteism \cite{hamby2018privacy}; South Asian women provided false information to protect themselves from online abuse \cite{sambasivan2019they}. \\
\hline
\textbf{Privacy-enhancing technologies (PETS)}\newline e.g., authentication, cloaking, encryption &\textbullet\ Social liability \newline\textbullet\ Erasure of records & Women in patriarchal societies used private modes and locks on devices, which may be seen as incriminating and invite coercion to obtain access \cite{sambasivan2018privacy}. People who are financially insecure who lose access to trusted devices lose access to services that require two-factor authentication \cite{sleeper2019tough}. \\
\hline
\textbf{Physical workarounds}\newline  e.g. hiding device, use of camera covers \& headphones &\textbullet\ Limits environmental awareness\newline \textbullet\ Vulnerable to physical coercion & People with visual impairments used headphones to avoid aural eavesdropping when using screen readers at the cost of physical safety \cite{ahmed2015privacy}. \\
\hline
\textbf{Asking for help} \newline  e.g. learning new practices, consulting network, websites, professionals &\textbullet\ Bad information \newline\textbullet\ Involves risk/trust\newline \textbullet\ Limited to help available & Professionals who provide support for survivors of intimate partner violence did not feel equipped to advise on identifying or coping with technology-enabled IPV \cite{freed2017digital}. People with visual impairments asked allies for help, but this risked trusting the ally with personal information \cite{hayes2019cooperative}. \\
\hline
\textbf{Collaborative \newline privacy practices}\newline  e.g. shared guidelines,  boundaries &\textbullet\ Loss of autonomy\newline \textbullet\ Involves risk/trust & LGBTQ+ adults considered not only their own privacy boundaries but also those of their families, ex-partners, and children \cite{blackwell2016lgbt}. Families co-developed privacy guidelines for shared devices in Bangladesh ~\cite{ahmed2017digital}. \\
\hline
\textbf{Third-party \newline protections}\newline  e.g. parents removing devices, organizations destroying info &\textbullet\ Loss of autonomy\newline\textbullet\ Outside of the person's control  & Art therapists removed identifying information from art created by persons with dementia to protect their privacy, but this also removed their voice \cite{cornejo2016vulnerability}. Canadian government's legal decision to destroy data documenting colonial abuses of indigenous people to protect their personal privacy also erased evidence of their abuse \cite{fullenwieder2018settler}. \\ \hline
\end{tabular}
\caption{Privacy Responses and Costs Framework: Types of responses to privacy threats observed in our dataset and their costs and consequences to marginalized people.}
\label{table:1}
\end{table}

\blue{
\subsection{Responses to Privacy Threats and their Costs and Consequences to Marginalized People}

Many existing privacy frameworks overlook the impact of marginalization on people's behaviors \cite{knijnenburg2022modern}. In table \ref{table:1}, we present a \textit{Privacy Responses and Costs} Framework that lists the ten main types of responses to privacy threats that we observed in the literature and the costs and consequences they carry for marginalized people. Research on the privacy calculus has established that people weigh costs and benefits as they make choices about privacy~\cite{dienlin2016extended}; our framework provides an account of costs and consequences involved in privacy decisions for marginalized groups. It is also noteworthy that weighing tradeoffs and making privacy decisions is challenging for people who are under stress \cite{leitao2019anticipating, sleeper2019tough, matthews2017stories}, adding a new dimension of complexity to the privacy calculus. Sannon and Cosley characterize privacy management as a costly form of invisible labor that has an outsized impact on marginalized groups or those without power \cite{sannon2019privacy}. In what follows, we discuss each of these privacy responses and their associated costs in turn.

\subsubsection{Apathy.} In some cases, people can feel helpless and may do nothing in the face of privacy and security threats, a phenomena McDonald describes as ``a sense of futility masquerading as apathy''~\cite{mcdonald2020politics}. For example, when exposed to online scams, people from economically disadvantaged communities may respond with feelings of resignation \cite{vitak2018knew}. Similarly, undocumented immigrants describe the government as an all-knowing entity whose surveillance is impossible to evade, leading to inaction \cite{guberek2018keeping}. The major cost associated with this response is continued exposure to the disproportionate harms faced by marginalized groups. 

\subsubsection{Non-use.} When faced with poor design and privacy threats, marginalized people may resist using technologies altogether, which can exacerbate the problem of exclusion. Economically disadvantaged populations in particular appeared to resist technology use in our dataset, which could limit their opportunities for jobs and social support~\cite{hui2018making, hamby2018privacy, sleeper2019tough}. For example, Vitak et al. described economically disadvantaged participants who were ``hesitant to use technology or outright shunned it, preferring to use analog methods for submitting applications, forms, and payments whenever possible—even when that decision carried additional financial costs or took longer,'' citing a participant who would not apply for jobs that required online applications \cite{vitak2018knew}.

\subsubsection{Withholding disclosure.} Self-censorship is a commonly discussed strategy for managing perceived risk, particularly of views that may give rise to interpersonal conflict~\cite{marwick2017nobody} or political speech that may have negative repercussions ~\cite{simko2018computer, jack2019privacy}. Marwick et al. observe that ``choosing to self-censor and limit one’s participation is a choice to be rendered invisible''~\cite{marwick2017nobody}. Self-censorship as a response to privacy threats can further silence marginalized voices. Moreover, withholding information may not be effective, with damaging consequences. Warner et al. describe how withholding HIV status can be ineffective, as people can also make inferences about undisclosed information \cite{warner2018privacy, warner2020evaluating}. Efforts to withhold disclosure may also be fruitless in the face of intentional adversarial threats like the use of stylometry to identify content creators \cite{forte2017privacy}. 

\subsubsection{Controlling disclosure.} Social media studies made up 26\% of the dataset, and a common privacy-protective behavior, especially on social media, centered around people making efforts to compartmentalize their identity or choose their self-presentation in ways that would allow them to pursue their goals while shielding them from risk. Some ways to do this are to use privacy controls to restrict access to one's information, to open multiple accounts on the same platform to keep various facets of one's identity separate~\cite{haimson2015online}, and to select platforms based on the degree of privacy they afford \cite{karusala2019privacy, devito2018too}. However, extensively locking down one's online profile also means that one's ability to express one's identity is restricted \cite{devito2018too}. The labor involved in controlling disclosure is also considerable. Fritz and Gonzales describe how one trans participant who was crowdfunding surgery ``took an entire day to go through all his Facebook connections and block approximately 600 people who were connected to his family so they would not see when he promoted his fundraiser on Facebook'' \cite{fritz2018privacy}. Pearce et al. describe the labor Azerbaijani young adults expend to segment audiences for political posts: ``those with two profiles engaged in a great deal of labor to manage the two---defriending people on one, adding them to the other, inventing innocuous reasons why a new profile was created, and so on'' \cite{pearce2018privacy}, and additionally describe the social cost of dissolving online ties to control disclosure. In a study of South African mobile privacy practices, Reichel et al. explain that using of privacy controls carried a financial cost associated with connectivity: ``Nearly every time a participant expressed awareness of these privacy settings, they followed by explaining they had an inability to actually access the privacy settings, often mentioning data costs'' \cite{reichel2020have}. 

\subsubsection{Privacy Lies.} Providing false personal information about oneself by telling ``privacy lies'' \cite{sannon2018privacy} appeared several times as a privacy strategy to deter threats. Sambasivan et al. describe the use of fake information as a strategy for South Asian women to protect themselves from abuse online \cite{sambasivan2019they}. Hamby et al. frame providing false information as a kind of privacy vigilanteism in rural Appalachia \cite{hamby2018privacy}. However, telling privacy lies requires cognitive effort to ensure the lies are not found out, and can also be risky, as being caught out in a lie can result in social or legal repercussions \cite{sannon2018privacy}.

\subsubsection{Privacy-enhancing technologies (PETS)}  
Papers in our dataset reported on a limited range of technological responses by marginalized groups to privacy threats. These largely involved private modes, encryption, and locking/adding additional authentication requirements. Despite the protections they offer, the use of PETs may incriminate the very people who need protection. Sambasivan et al. report in a study of women in India, Pakistan, and Bangladesh that ``private
modes are often associated with ‘secret’ activities, threatening participants’ values of openness as they performed culturally appropriate gender roles'' \cite{sambasivan2018privacy}. Similarly Ahmed et al. observed in a study of shared mobile phone use in Bangladesh that ``almost half of our participants reported that locking specific data or applications might also raise suspicions in the mind of their partners''~\cite{ahmed2017digital}. Locking down devices and applications may also result in further coercion or physical harm in the context of intimate partner violence \cite{leitao2019anticipating, freed2018stalker}. Because of this cost, hiding the fact that privacy protections exist was surfaced as a design recommendation in multiple papers. Naseem et al. quoted a participant in Pakistan who applauded the design of secret PETs, ``That way, at least one won’t come across as suspicious, especially
since men in our society are very distrustful and suspicious''~\cite{naseem2020designing}. However, leaving no trace of activity carries additional consequences for people in abusive relationships in that it eliminates documentation of abusive behaviors~\cite{ramirez2019communication}. Additionally, the use of two-factor authentication can contribute to a victim's powerlessness if the victim loses access to devices required to authenticate, possibly due to interference by the abuser \cite{sleeper2019tough}. 

\subsubsection{Physical workarounds.}  
Hiding devices, covering cameras, using headphones to prevent others from overhearing screenreaders--these are all documented physical workarounds that members of marginalized groups have to privacy threats. Ahmed notes that many visually impaired participants used headphones to ensure privacy when using screenreaders; however, ``since visually impaired people rely on hearing in order to sense the environment, headphones leave them more vulnerable to other privacy and safety concerns'' \cite{ahmed2015privacy}; thus the headphones might protect their information privacy at a cost to their physical safety. People in abusive relationships may also physically hide their devices to avoid unwanted snooping~\cite{matthews2017stories}, but hiding devices comes with many of the same threats of physical or emotional coercion described in the above section on PETs. 

\subsubsection{Asking for help.}  
Asking for help may involve learning new privacy practices from social contacts, or consulting privacy resources and professionals. When people rely on their social network, a practice found to be more common among lower SES individuals~\cite{redmiles2017digital}, the help they receive is only as good as the knowledge in their network. Poor advice can carry a high cost for those in vulnerable positions and even trusted professionals may not be sure what to advise. For example, professionals who support survivors of intimate partner violence don't always know what to advise~\cite{leitao2019anticipating}. 
Receiving help from others can also entail divulging sensitive information to third parties, which requires trust and can introduce a new privacy risk~\cite{vitak2018knew, mentis2020illusion, simko2018computer}.

\subsubsection{Collaborative privacy practices.} Collaborative practices sometimes arise as a response to concerns about privacy threats within family units or other relationships. Practices like establishing shared rules and boundaries ~\cite{ahmed2017digital} provide a social alternative to PETs like app locks and encryption, but require high levels of trust and may compromise autonomous decision-making.

\subsubsection{Third-party protections}
In literature about marginalized populations with heightened vulnerabilities, it is unsurprising to find examples of third parties with power making decisions that impact marginalized groups. Examples in our dataset range from husbands making decisions for wives in patriarchal societies \cite{karusala2019privacy}, governments taking action on behalf of vulnerable groups \cite{fullenwieder2018settler}, and families and therapists taking responsibility for adults with cognitive decline~\cite{cornejo2016vulnerability, muller2010dealing}. Costs of third-party protections include a loss of autonomy and control; when others make privacy decisions, the line between helpful and paternalistic can be difficult to see. Mentis et al. explore this tension in work with adults with cognitive decline and their partners \cite{mentis2020illusion}; they find that although partners aim to negotiate security decisions, the reality is often that decision-making ends up being one-sided by the caregiver partner. 

}

\subsection{High-level privacy-related tensions and trade-offs for marginalized technology users.}

In addition to the granular costs and consequences described above, we identified several high-level privacy-related tensions in studies of marginalized groups' technology use:

\subsubsection{Privacy vs. disclosure of identity} Most uses of technology entail some degree of identity disclosure, ranging from highly visible disclosures like creating a personal account with legal identifiers, to less visible disclosures like inadvertently sharing location via network data. For marginalized groups, the privacy threats associated with everyday use may be acute, even if the groups themselves are not cognizant of the threats~\cite{guberek2018keeping}. In some cases, technologies create unique avenues for sharing among marginalized groups that come with elevated threats. For example, LGBTQ social media users may use platforms to connect and explore their identity~\cite{haimson2020trans} but also risk stigma from unintended audiences if their privacy strategies are compromised~\cite{devito2018too}. Marginalized groups engaging in activism~\cite{lerner2020privacy} or participating in online collaborations~\cite{forte2017privacy} may disclose identity characteristics by virtue of these activities. In other cases, privacy breaches can be the result of secondary data use, for example by analyzing meetup data to infer LGBTQ identity~\cite{chung2017privacy} or social media posts to infer mental health status~\cite{weerasinghe2019because}. 

\subsubsection{Privacy vs. support} Because marginalized groups by definition experience some form of stress, support seeking is particularly salient. Risking disclosure of vulnerabilities to receive support is not unique to technology-mediated contexts (e.g.~\cite{guberek2018keeping}), but technologies can exacerbate the privacy threats associated with support seeking.~\citet{srinivasan2018privacy} observe in their research on identity infrastructures in India that low income and marginalized people often ``are the people who most need benefits from the state, and to receive benefits, they must identify themselves,'' which can introduce risk of stigma and persecution. In a different context, strikingly similar trade-offs arise: for trans men who crowdfund financial support for top surgery, support-seeking can entail highly public disclosures~\cite{gonzales2017prioritizing, fritz2018privacy}. Sambasivan et al. investigated the use of a phone messaging system intended to help deploy social and health services to urban sex workers who are vulnerable to physical violence and health issues, but for whom receiving such support could mean stigma or further physical threat~\cite{sambasivan2011designing}.

\subsubsection{Privacy vs. autonomy} Privacy is an important way of protecting individuals' freedoms (see  ~\cite{westin1968privacy}). As such, oversight of others' location, well-being, or activities is often paternalistic or invasive, but some of the research in our dataset discussed scenarios in which forms of surveillance were used to facilitate autonomy. One example of the challenges of navigating privacy and autonomy arises in the work of victim service providers (VSPs). These organizations support survivors of human trafficking. In order to help ensure that survivors of human trafficking are not revictimized and in a direct bid to protect survivors' freedoms, VSP shelters may surveil communications, particularly of minors, to enforce rules and monitor for risky behaviors~\cite{chen2019computer}. Similarly, Mentis et al. identified threats, tradeoffs, and design considerations with respect to privacy of older adults with cognitive decline and their caretakers~\cite{mentis2019upside}; they note that while collaboration between adults with cognitive decline and caregivers is critical, supporting collaboration entails privacy concessions that can create opportunities for exploitation. Similarly, location tracking can be viewed as a safety measure that supports adults with dementia in retaining more freedom~\cite{muller2010dealing}, but it can also be subject to potential abuse~\cite{dahl2012there}. This handful of papers highlight important boundaries where social norms and cultural standards around the limits of privacy and acceptable privacy trade-offs for safety and wellbeing are negotiated.

\subsubsection{Individual vs. collective} Although privacy theorists often weigh the interests of the individual against those of states and organizations (see~\cite{westin1968privacy,brandeis1890right} for foundational examples in Western culture), in some cases, privacy concerns are collective. That is, more than one person may work together to protect the shared privacy interests of a group. The literature on marginalization and privacy included several examples of such framing. For example, LGBTQ+ parents on social media find that their privacy depends not only on their own disclosure decisions but also the disclosures and behaviors of their network. Further, their self-disclosures can also impact the privacy boundaries of those who make up their network, such as their children, partners, and former partners, and potentially expose the network to stigmatization as well \cite{blackwell2016lgbt}. In some cases, additional help may be needed in navigating privacy---for example, in the context of patients with dementia, Cornejo et al. describe how negotiating privacy is a process that is shared between the patients, family members, and care professionals \cite{cornejo2016vulnerability}. People's privacy preferences may also be ``socially-negotiated'' rather than purely their own, as with people managing bipolar disorder whose family members and care team who would like them to regularly share their personal data in the form of a ``check-in'' \cite{petelka2020being}.

\subsection{Technologies and the way they handle privacy can either contribute to or impede the process of marginalization.}

We identified two common narratives across the literature that were connected to the kind of mediating role that technologies can have as the relationship between privacy and marginalization plays out. On one hand, technologies can be viewed as a mechanism for greater equity and freedom, slowing or perhaps optimistically even reversing the process of marginalization for certain groups, such as technologies that provide assistive support~\cite{wang2019eartouch} or safe spaces for disclosure and support seeking (e.g., ~\cite{hong2012designing, carrasco2018queer, naseem2020designing}). 

On the other hand, technology can be viewed as a mechanism that supports and strengthens processes of marginalization by disproportionately introducing privacy threats and harms or furthering exclusion. In some cases, the design of technologies is insufficient to protect the privacy interests of marginalized groups. For example, insufficient control over disclosures of  HIV-positive status in dating apps can lead to ``privacy unraveling''---particularly for men who have sex with men---which can further exacerbate their marginalized status~\cite{warner2018privacy}. Being from a marginalized group means that technologies may be designed in ways that exclude people from their use at all, as Rennie et al. explain, ``social dynamics and obligations can prevent Aboriginal people from using devices and settings in the way they were intended''~\cite{rennie2018privacy}. Being prevented from using technology by virtue of poor design is yet another form of exclusion that marginalized groups experience. In some cases, poor design or the threat of privacy breaches may lead to non-use or resistance~\cite{hamby2018privacy} which, while sometimes framed as a form of agency and empowerment, can also result in exclusion from digital public spaces~\cite{forte2017privacy}. Reichel goes further to examine how the very concept of privacy as a goal is embedded within and therefore yields social and technological structures that exclude marginalized groups~\cite{reichel2017race} and~\citet{gangadharan2017downside} notes that inclusion efforts intended to assist marginalized groups may expose them to greater threats of surveillance and risk. 

\section{What are key themes in the papers' recommendations?} 

We organized the recommendations in the papers into three broad categories---conceptual, technological, and behavioral recommendations---that we discuss next.

\subsection{Conceptual recommendations}

\subsubsection{Prioritizing autonomy and dignity in design}

Several authors recommend rethinking the ways we talk about privacy and the concepts we use to understand and design privacy for marginalized groups.  Akter et al. call for ``humanizing'' assistive technologies specifically because they found that ``camera-based assistive systems were creating a lack of security in people's daily lives---that is, these systems were serving to further marginalize their identities''~\cite{akter2020uncomfortable}. Reichel rejects the concept of privacy, seeing it as inherently flawed in that it requires identification of a threatening ``other,'' and advocates instead to organize what is now privacy discourse around the more fraternal concept of dignity~\cite{reichel2017race}. ~\citet{fullenwieder2018settler} similarly critique the notion of privacy as an individual right in their analysis of how privacy was leveraged to support the destruction of testimony about state-sanctioned violence against indigenous populations in Canada.   

Designing technologies that do not threaten the privacy of marginalized groups also involves assessing which values are given importance in the design process, and considering how these values might impact marginalized groups. Values can be in conflict: for example, remote monitoring systems for people with dementia can prioritize the value of keeping them safe over preserving their privacy. In light of such conflicts, Dahl and Holbø recommend that value elicitation should be a necessary part of designing in such settings to evaluate technological biases and impacts \cite{dahl2012there}. Similarly, Wan et al. argue that flexibility should be built into technologies so that they can be adapted to fit potentially diverse needs in different organizational and family contexts \cite{wan2014addressing}. 

Prioritizing people's agency is one way of ensuring technologies do not harm marginalized groups. For example, automatic gender recognition software can engender numerous harms for transgender people; Hamidi et al. caution designers to consider whether gendering users is truly necessary and to exclude gender from their design if possible. In cases where such technologies are used, they recommend providing users with the agency to opt out of being gendered \cite{hamidi2018gender}. 

\subsubsection{Recognizing the influence of power relations on technology use}

Studying marginalization means studying power. As part of their focus on marginalized groups, many researchers call for designers ``to consider the ways that technologies may not be one-size-fits-all'' \cite{fernandez2019don}. These researchers highlight the need for designers to pay more attention to how power relations structure the ways marginalized groups use technology and to tailor technologies accordingly. For example, locking a mobile phone may not be effective in a heavily patriarchal context where a woman can be coerced by her husband to unlock it \cite{ahmed2017digital}. Similar dynamics may occur due to familial and sociocultural power relations, especially in contexts where it is the norm for multiple people to share the same device \cite{sambasivan2011designing}, such as in South Asia \cite{sambasivan2019they}. Conducting digital risk assessments for technologies would also identify privacy risks and gaps in privacy protection based on different use cases, such as assessing if certain technologies pose risks to survivors of intimate partner violence, given their unique threat model where they are avoiding a known other \cite{freed2017digital}.

\subsection{Technological recommendations}

Most recommendations, particularly in HCI, centered around how technologies can be designed to better address the unique privacy needs and concerns of marginalized groups.

\subsubsection{Providing greater control over information}

Since many marginalized populations can face heightened risks from having their privacy violated, many studies stressed the need to afford people with greater control over their information (e.g., \cite{haimson2015online, gonzales2017prioritizing, carrasco2018queer, fernandez2019don}). To this end, a common design suggestion was that technologies should have granular privacy settings so that users can better control the visibility of their content, thereby avoiding the risks of context collapse. For example, Carrasco et al. discuss how LGBTQ+ social media users practice ``selective visibility'' by being more ``out'' to LGBTQ+ audiences but not cisgender or heterosexual audiences; they suggest that social media platforms that facilitate this form of selective sharing---for example, through supporting the use of multiple profiles---would give marginalized social media users more agency over their self-presentation (and by extension, their privacy and safety) \cite{carrasco2018queer}. 

Designing technologies that do not diminish the privacy of marginalized groups also requires designers to consider the ``labor and risk involved in conveying sensitive information about the self to others'': for example, some groups---such as trans individuals---may not want to seen by the broadest audience possible when using online dating platforms, as this could open them up to harassment, and instead would benefit from controlling the visibility of their profiles \cite{fernandez2019don}. %p39

\subsubsection{Facilitating management of communal and networked aspects of privacy}

Several papers stressed the need for designers to consider communal aspects of privacy. In addition to managing their personal privacy boundaries, people often have to navigate collective boundaries as well, and granular privacy controls on online platforms would help such networked privacy management \cite{blackwell2016lgbt}. In cases where multiple stakeholders jointly negotiate privacy, such as in the case of people with dementia and caregivers, systems could support cooperatively setting privacy preferences \cite{cornejo2016vulnerability}.

People can also face privacy threats from others who have access to their devices, either with or without their consent. In such cases, a common design recommendation is to facilitate secrecy and the hiding of sensitive information. For example, mobile phones could be set up to allow an individual to hold multiple accounts that are kept secret if others access the device \cite{ahmed2017digital}. Similarly, enabling multiple users to create individual accounts on shared devices also helps maintain each individual's privacy \cite{leitao2019anticipating}. Devices could also enable on-demand information ``hiding'', a feature that would be especially useful for groups who could face severe consequences if their information were accessed by the wrong parties, as in the case of undocumented Latinx immigrants \cite{guberek2018keeping}.

\subsubsection{Making privacy decisions easier}

Another type of recommendation centered around making it easier for marginalized groups---and people in general---to make informed privacy-related decisions. One way of doing this is to improve the usability of privacy and security features and settings, including when people are under dire stress, as in the case of avoiding intimate partner abuse \cite{matthews2017stories}. Another option is to design for transparency, so that they have more clarity around when their information is being collected and how it might be used. Sometimes this transparency is important with respect to how data will be used by social media companies and service providers~\cite{cabanas2018unveiling}, but also with respect to how information is shared with other parties. For example, in the context of an app that helps people with bipolar disorder continuously disclose mental health information to trusted others, managing this continuous sharing in a sensitive way requires interfaces to clearly indicate who can see what data at any given time ~\cite{petelka2020being}.

\subsubsection{Building in technical safeguards}

Whereas many of the above-discussed technical recommendations hinged on user experience and interface features, several papers also discussed how technical infrastructures could be better designed to safeguard marginalized populations. For example, in the call for humanizing assistive technologies, ~\citet{akter2020uncomfortable} note that computer vision algorithms must be designed to detect not only objects, but features of context that matter to the humans who use them. Systems that provide infrastructure for anonymous online activity can protect people with marginalized identities in a variety of contexts, for example, survivors of abuse who wish to report abusers~\cite{chen2019computer}.

\subsection{Behavioral recommendations} 

Some researchers stressed the need to help people engage in behaviors that keep them safe and align with their privacy needs. One way to do this is to provide people with education and resources. Access to privacy-related education can be particularly useful for groups such as undocumented immigrants, who may not be aware of the intricacies of the technological privacy and security threats they face \cite{guberek2018keeping}. Educating people about how to protect their privacy can help them continue to use technologies while mitigating the risks in doing so, as in the case of contributors to open collaboration projects like Wikipedia who have marginalized attributes \cite{forte2017privacy}. In both these studies, the authors stress the need for such social interventions to occur in conjunction with technological solutions that focus on improving the technologies themselves.

In some cases, marginalized groups may be well aware of the fact that they face privacy and security threats, but feel limited in their ability to address these threats. Research on intimate partner abuse suggests that survivors need targeted instructional materials that will help them avoid their abusers, such as information on how to use security features like two-factor authentication \cite{matthews2017stories}. In response to this need for access to resources, two studies in our dataset explored the usefulness of providing survivors of intimate partner violence with security consultations with a trained technologist, finding that these consultations were generally perceived as valuable and also uncovered security vulnerabilities that participants were not aware of \cite{freed2019my, havron2019clinical}. 
            
Despite multiple studies pointing to a need for improved access to privacy education, a challenge in improving privacy education among marginalized groups is that digital literacy programs are not always well-attended; in response to this challenge, Vitak et al. point out that stories are often an effective means of spreading information in low-income communities and could be a way of sharing privacy-related knowledge and resources \cite{vitak2018knew}. Reichel et al. suggest that ``lightweight privacy on-boarding interfaces'' could help resource-constrained users, including making privacy settings available offline for those who have limited connectivity \cite{reichel2020have}. People may also be able to improve their digital privacy skills if they have unrestricted, private access to the Internet, but many disadvantaged communities rely on libraries and schools to access the Internet where their usage is time-limited; thus, addressing the digital divide remains crucial to increasing both the autonomy and privacy literacy of marginalized groups \cite{li2018privacy}. 

\section{Where do we go next? Charting Future Directions}

Our initial dataset included 2,823 privacy-related papers that were published between 2010 and 2020; of these, only 3\% (88) focused on marginalized contexts. Although our dataset does not include all venues where such work might appear, this highlights how profoundly understudied this area of research remains. We also found that research in this area is growing over time, and we end this paper by discussing some potential avenues for future research in terms of focus, topic area, methods, and research practices.

\subsection{Broaden How We Problematize Privacy and Marginalization} 

Although we identified several common themes that cut across studies, papers in our dataset did not clearly converge on a shared articulation of problems or solutions. We note that, although all the papers examined marginalized groups that are, by definition, marginalized as a function of social norms and structural inequities, only a handful of papers problematized privacy in the context of social structures, policies and laws. We are confident that nearly all researchers doing work on privacy at the margins understand that technologies, policies, and social structures intersect and inform one another; however, efforts to examine these intersections were rarely central features of papers. In contrast, empirical studies, which were the most common type of contribution in our dataset, tended to focus on people's experiences and behavior. These studies mark an important shift towards representing the voices of users who have been traditionally excluded from technology research. Alongside these much-needed contributions, we suggest that researchers leverage and build on critical lenses that highlight structural and systemic aspects of marginalization that underlie technology design and use. Correcting inequities in privacy is also a problem with potential policy implications. Our dataset yielded few if any policy recommendations, and we see a need for more interdisciplinary work that bridges technology design and policy. 

\subsection{Fill Understudied Research Gaps}

While research on marginalization and privacy is growing rapidly, particularly in the past three years, this growth has not been uniform across research areas. As we coded our dataset of existing literature, it was quickly apparent that some research topics at the intersection of privacy and marginalization remain undeveloped. 

The paucity of discussion around race and privacy in our dataset was conspicuous. Papers that examined issues of race and power as their central focus were few, and mainly within the Communication dataset \cite{reichel2017race, fullenwieder2018settler}. Additionally, some papers looked at contexts in which race was inextricable from the context under study, such as refugees and other immigrants, but for the most part, these did not explicitly reflect on the influence of race in the experience of marginalization. Additionally, race was sometimes mentioned as a factor to consider in contexts like crime prevention or low-income neighborhoods, but was not used as a lens through which to understand privacy concerns or as an analytical tool in empirical work. This is surprising for HCI in particular, given a trend toward considering technologies as potential instruments of oppression and marginalization and increasing attention to conceptual frames that address features of race, like critical race theory ~\cite{ogbonnaya2020critical} and---at times controversially---intersectionality~\cite{rankin2019straighten}. Although such frames were at times mentioned in discussion, they were generally absent from the body of empirical work we reviewed. To fill this important gap in the literature, we see a dire need for privacy research---particularly within HCI and Privacy-focused venues---that includes and centers race. 

It is worth noting that this is a new subfield; even the most common topics are still relatively understudied and remain rich areas for further exploration. This is also true within various topic areas. For example, disability and LGBTQ+ issues were the two most common topics and together made up almost half of our dataset, but certain subpopulations were less studied than others, such as people with mental health conditions (n = 3) and queer women (n = 1), respectively. As such, our review illustrates the current breadth of work in this space but also shows that there is much room for future work across topics. 

\blue{In order to fill these gaps, we see an immediate opportunity for conferences and journals to foster research in this area. Scholarly leadership can support these efforts by mentioning topics related to marginalization in CfPs, dedicating special issues, panels, and workshops, identifying keynotes and themes. Moreover, program committee members and reviewers must recognize the value of studying marginalized groups independent of comparisons to or in the interests of designing for the majority population.}

\subsection{Diversify Methods and Questions}

Qualitative methods---particularly semi-structured interviews---were by far the most common methods used by researchers. This makes sense given that the goal of much empirical work thus far has been to understand the privacy-related needs and experiences of marginalized user groups. While these methods are likely common because they are the most appropriate for the types of questions the community has been asking, our findings indicate the potential to leverage other methodologies to complement and build on existing research questions. Within the umbrella of qualitative research, ethnographies, textual analyses, and case studies are all examples of methods that are under-represented in this space but may still be well-suited to the questions being posed. There is also potential to leverage quantitative methods to a greater degree; for example, to measure privacy inequities and disparities, or to experimentally test technology designs intended to mitigate these disparities. Further, a minority of papers used participatory methods to co-design privacy-aware technologies with users, and the privacy community may benefit from greater use of these methods, particularly as mechanisms by which to involve marginalized people in design. While it is important for researchers to acknowledge that new tools may not be wanted or needed by the individuals they intend to serve, the few participatory studies in our dataset suggest that there is potential to work in conjunction with marginalized communities to co-develop tools or re-design existing technologies to better meet their needs and practices. To do so, we can also look to HCI research on other marginalized contexts, such as participatory action research in under-resourced neighborhoods \cite{hui2020community}, for guidance.

\subsection{Develop Shared Best Practices}

We found wide variation in both research and reporting practices across studies, such as the amount of detail provided in methods sections and decisions around ethical considerations, such as compensation. Particularly in the context of studying marginalization, we see an urgent need to consider, justify, and report study details, ranging from why a given recruitment strategy was chosen to potential harms from the research and how these were mitigated. These are challenging aspects of study design, and reporting the decision process underlying these research choices is important for research communities to engage in discussion/critique and to develop shared practices and ethical standards. 

Here, we discuss some research considerations that researchers need to reflect on when crafting and conducting our studies and report on when publishing our papers as a first step towards developing shared best practices for studying privacy and marginalization. 

\subsubsection{Plan for and reduce a wide range of potential harms}

Reducing harm and maximizing benefits are central tenets of  international standards for ethical research~\cite{code1949nuremberg,belmont}. Categories of ``vulnerable'' populations highlighted by bodies like Institutional Review Boards (IRBs)---for example, prisoners, pregnant women, children---do not cover all possible at-risk groups. Marginalization by definition creates vulnerabilities. When working with marginalized groups, researchers need to take particular care to understand the potential risks involved in their participation. Whereas a large portion of our data corpus was focused on identifying privacy risks, these papers rarely discussed the privacy risks and precautions related to study design beyond anonymizing data or using pseudonyms. According to a review of anonymization practices at CHI, this is not atypical \cite{abbott2019local}. Moreover, some study designs in our corpus may have introduced novel risks to the population, for example through automated detection of marginalized identity features, without a robust discussion of the ethical implications of such research endeavors. As a research community it is important not only to carefully consider, but also unambiguously report our considerations of  potential harms or unintended consequences and how these risks are mitigated in study designs. 

Another important consideration to reduce harms in research is in the language that researchers choose to represent marginalized people and experiences. Language that others, diminishes, or inadvertently stigmatizes marginalized groups can introduce harms through the research itself. Researchers should take care to learn and discuss the preferences of the groups involved in research and report if and how specific choices were made around labels and terminology where controversy or disagreement exist.  

Other strategies used to mitigate harm in our dataset included having a therapist on hand to assist if interview participants experienced distress \cite{leitao2019anticipating}, along with other strategies already discussed in Section 6.3. Researchers who conduct interviews on sensitive topics or with groups facing adversity may also find Kasket's \cite{kasket2009protocol} protocol for responding to participant distress useful.

Finally, while studying marginalization, it is also worth considering potential harms to researchers themselves. Researchers can face burnout as well as physical safety risks when conducting research on sensitive topics \cite{ensign2003ethical}. A handful of papers in our dataset discussed the steps that were taken to ensure researcher well-being, and these discussions were useful to understand some of the challenges that may surface when conducting research in this space, as well as how these risks might impact the research process and the scope of the findings. For example, a study in our dataset that involved conducting interviews in unsafe neighborhoods noted that the interviews were strictly time-bound to ensure the safety of the researchers, sometimes at the cost of asking additional follow-up questions \cite{reichel2020have}.

\subsubsection{Report compensation}

A review of HCI papers published by the ACM found that the majority of user studies did not report whether participants were compensated \cite{pater2021standardizing}. Our dataset suggests that researchers working at the intersection of privacy and marginalization may be more likely to report this, but still, about half of the user studies in our dataset did not include this information. 

We echo Pater et al.'s call for researchers to report this information, including their rationales. When providing compensation, researchers could consider whether there are implicit biases in their compensation choices---for example, not all participants may want an Amazon gift card \cite{pater2021standardizing}, and in the context of studying marginalized groups, there may be differences in access that make some forms of compensation more appropriate or desirable than others. A few studies in our dataset described consulting with community members or partners to decide on appropriate compensation, which is a practice that could be particularly well suited for researchers in this space to adopt.

\subsubsection{Discuss positionality}

Positionality is an important feature of research---researchers' identities, beliefs, experiences, and backgrounds directly inform their selection of methods, engagement with participants in the case of human-subjects research, and interpretations of data. How to go about reporting positionality is a complex decision---for example, positionality statements that involve identity self-disclosures may disproportionately harm marginalized researchers. We do not make prescriptive statements about whether all studies should explicitly include this information or not. However, considering positionality is a vital part of the research process, particularly in the context of working with marginalized groups. In our dataset, a minority of papers reported positionality, often in the form of disclosing researchers' own marginalized identities. We suggest, echoing~\citet{liang2021embracing}, that positionality statements need not always include identity characteristics. Positionality statements do not confer legitimacy, and although they may provide useful information about researchers' insider status within a community of study, the goal of a positionality statement is to provide context that helps readers understand the research being presented. In some cases, researchers' political beliefs, epistemological commitments, and disciplinary training may be more helpful in understanding how research was conducted than specific identity characteristics. For example, in the case of this paper, both authors have experienced marginalization in multiple dimensions, which may help readers contextualize our interest in and commitment to this topic, and we are both privacy researchers who approach research from a critical, interpretivist perspective. These are important things to know about us, since we are offering research advice, whereas specific disclosures around our race, sexual orientation, and other identity features may not be.

\subsection{Limitations and Future Work}

\blue{
Literature reviews have limitations that revolve around the selection of conferences and databases, as well as the criteria used to search for and filter papers. 

%VENUES/CONFERENCES
First, our sample was scoped to a large but still limited set of venues. This is because our sampling strategy was designed to uncover themes in a broad, interdisciplinary set of papers at the intersection of privacy and marginalization, rather than to identify a comprehensive dataset of all papers published in this space across fields. As a result, we limited our search to SIGCHI-sponsored venues, select Communication journals, and select venues that focus on privacy and security research. While our exploratory searches suggested that these were some of the most common spaces for research relevant to our focus, it also constrained our sample. We think there is potential for future work to examine other conferences and fields in greater detail, including venues that focus on historically marginalized groups, such as the ACM Conference on Computers and Accessibility (ASSETS). 
%KEYWORDS

Second, the criteria we used to search for and filter papers also influenced our sample. When searching for relevant papers in ACM SIGCHI-sponsored venues and Communication journals, we constrained our search to just one keyword (``privacy''). This allowed us to limit the size of our dataset and remain consistent across subfields; however, in doing so, we may have missed work on marginalization as it intersects with other topics related to privacy, such as trust and disclosure. Since we filtered papers based on whether they contained the search term ``privacy'' in the title, abstract, and/or keywords, we likely missed some relevant papers that are not focused on privacy but nevertheless have interesting privacy-related findings pertinent to our focus.

%SUBJECTIVE - what is marg, populations
Further, ``marginalization'' is a term with blurry boundaries and multiple definitions that are open to interpretation. Manually filtering our dataset to identify papers on marginalization required us to draw on several definitions of marginalization that often differed from each other, as well as our conceptualizations that were necessarily constrained by our own readings and positionalities. Our initial step was to read widely about how the concept of marginalization has been developed and used in a variety of fields, and we engaged in ongoing discussions about marginalization within our research group. Ultimately, our decision about whether to include studies in our dataset was informed by our understanding of what different social groups experience, our interpretations of definitions, and colored by our own beliefs and backgrounds. While we chose to focus on the experiences of marginalized adults, we also see potential for future work to explore research on children, who are a vulnerable population with unique privacy risks.
}

\section{Conclusion}

The goal of this paper was to serve as a roadmap for current and prospective researchers working at the intersection of privacy and marginalization by providing a review of current knowledge in the field, the practices through which it has been generated, and to chart a way forward. Our review of 88 papers published between 2010--2020 in HCI, Communication, and Privacy-focused venues found that this is a fast-growing area of study with wide variation in topics and research practices. \blue{Many existing privacy frameworks do not account for marginalization \cite{knijnenburg2022modern}, and in response, we introduced the \textit{Privacy Responses and Costs} framework to reflect the range of privacy responses people enact and the costs and consequences of these responses to marginalized groups.}

We also uncovered topics that need further study, such as race, the structural aspects of marginalization, and the role of policy, as well as the potential to use more diverse methods in our practices, including quantitative and participatory methods. Finally, given our focus on marginalized groups, we see a need to discuss and report research practices in greater detail, particularly around the ethical considerations of our work, and we put forth some suggestions for establishing shared best practices to do so. 

\begin{acks}
The authors thank Mako Hill, Rachel Greenstadt, and Kaylea Champion for their feedback throughout this study. The work was supported by the National Science Foundation (Award \#2031951). The first author was also supported by the NSF under Grant \#2127309 to the Computing Research Association for the CIFellows Project. The findings and recommendations in this paper are those of the authors and do not necessarily reflect the views of the NSF or the CRA.
\end{acks}

\bibliographystyle{ACM-Reference-Format}
\bibliography{SLR, dataset_papers}

\end{document}